\documentclass{pnastwo}

\usepackage[pdftex]{graphicx}
\usepackage[final]{pdfpages}

\usepackage{amssymb,amsfonts,amsmath}

\contributor{Submitted to Proceedings
of the National Academy of Sciences of the United States of America}
\begin{document}

\title{Landau spectrum and twin boundaries of  bismuth in the extreme quantum limit}
\author{Zengwei Zhu\affil{1}{Laboratoire Photons Et Mati\`ere (UPMC-CNRS), ESPCI, 75005 Paris, France},
Beno\^it Fauqu\'e\affil{1}{}
Liam Malone\affil{2}{Laboratoire National des Champs Magn\'{e}tiques Intenses (CNRS), 38042 Grenoble , France},
Arlei B. Antunes\affil{2}{},
Yuki Fuseya\affil{3} {Department of Materials Engineering Science, Osaka University, Toyonaka, Osaka 560-8531, Japan}
and Kamran Behnia\affil{1}{}
}
\contributor{Submitted to Proceedings of the National Academy of Sciences
of the United States of America}

\maketitle

\begin{article}
\begin{abstract} The Landau spectrum of bismuth is complex and includes many angle-dependent lines in the extreme quantum limit. The adequacy of single-particle theory to describe this spectrum in detail has been an open issue. Here, we present a study of angle-resolved Nernst effect in bismuth, which maps the angle-resolved Landau spectrum for the entire solid angle up to 28 T. The experimental map is in good agreement with the results of a theoretical model with parabolic dispersion for holes and an extended Dirac Hamiltonian for electrons. The angular dependence of additional lines in the Landau spectrum allows to uncover the mystery of their origin. They correspond to the lines expected for the hole Landau levels in a secondary crystal tilted by 108 degrees, the angle between twinned crystals in bismuth. According to our results, the electron reservoirs of the two identical tilted crystals have different chemical potentials and carriers across the twin boundary have different concentrations. An exceptional feature of this junction is that it separates two electron-hole compensated reservoirs. The link between this edge singularity and the states wrapping a three-dimensional electron gas in the quantum limit emerges as an outstanding open question.
\end{abstract}


\section{Introduction}

\dropcap{S}emi-metallic bismuth is a fascinating solid. Its average bulk diamagnetism is larger than in any non-superconducting solid, its  magnetoresistance is huge and it dwarfs any other solid with its Nernst coefficient. During the last century, the electronic properties of bismuth were subject to numerous studies (See \cite{edelman,issi,dresselhaus} for reviews). In particular, studies of de Haas-van Alphen effect and Shubnikov- de Haas effects led to a precise determination of the components of the Fermi surface\cite{smith,bhargava}, which consists of a hole ellipsoid at T-point and three identical electron ellipsoids at L-point\cite{liu}.

Recently, there has been a renewal of interest in the electronic spectrum of bismuth at very high magnetic fields in the vicinity of the quantum limit\cite{behnia1,luli,fauque1,fauque2,yang}. The quantum limit is attained when the magnetic field is strong enough to confine electrons to their lowest Landau level(s). With a magnetic field of affordable magnitude, this limit can only be attained in low-density metals like bismuth or graphite.  Emergence of collective effects in a three-dimensional electron gas in such a context have been a subject of theoretical speculation\cite{halperin,macdonald}.

The measurements of the Nernst response in bismuth up to a magnetic field of 12 T led to the observation of giant quantum oscillations \cite{behnia2}. The signal was found to sharply peak at each intersection of a Landau level and the chemical potential. Similar oscillations were also found in graphite\cite{zhu1}, another bulk semi-metal leading to several theoretical studies offering explanations for the occurrence of these giant Nernst quantum oscillations\cite{bergman,sharlai2, luk}.

The extension of the Nernst data up to 30 T led to the observation of additional Nernst peaks at magnetic fields exceeding the quantum limit\cite{behnia1}. These peaks were unexpected and, given the known structure of the Fermi surface in bismuth, could not be attributed to any Landau level. Moreover, angle-resolved torque magnetometry measurements found a complex phase diagram for a magnetic field slightly tilted off the trigonal axis\cite{luli}. These experimental findings raised the issue of collective effects in bismuth at high magnetic fields. In this context, the high-field phase diagram of bismuth became subject to a theoretical reinvestigation\cite{alicea,sharlai}. At high magnetic fields, the chemical potential continuously shifts to preserve charge neutrality. Therefore, quantum oscillations are no longer a periodical function of $B^{-1}$.  Further complexity arises due to an angle-dependent Zeeman energy and cyclotron mass for both electrons and holes \cite{smith}. Very recently, a fair agreement was found between the results of a theoretical model treating the electrons with an extended Dirac Hamiltonian and the experimentally-detected Landau spectrum up to 12 T\cite{zhu2}. The results narrowed down the possible choice of theoretical parameters to describe the system in presence of moderate magnetic fields but the origin of the additional lines remained an open question.

Several explanations for the origin of additional Nernst peaks have been proposed. It was suggested that they had been generated by misalignment\cite{sharlai}. But this was ruled out by measurements of the Nernst effect in presence of a tilted magnetic field \cite{yang}. It was speculated that surface states could give rise to these additional peaks\cite{seradjeh}. But both the amplitude and the profile of the Nernst peaks point to a bulk origin\cite{behnia3}. Moreover, their complex angle dependence does not follow what is expected for a 2D state. Until now, the origin of these additional lines on top of the complex Landau spectrum has remained a mystery.

Previous angle-resolved Nernst measurements mapped the Landau spectrum in a limited angular window\cite{yang}. Here, we present for the first time a map of Landau spectrum extended up to 28 T and across the whole solid angle. Comparing the experimental map with the results of the theoretical model for bulk bismuth allows us to find a solution to this mystery. All lines resolved by the experiment match the theoretical expectations. In particular, the additional lines follow an angular variation corresponding to a crystal tilted by 108 degrees as expected in case of a twinned bismuth crystal. The results suggest that the entire phase diagram of bismuth, at least up to 28 T, can be explained within the single-particle theory. Surprisingly, however, we find that the agreement between theory and experiment can be attained only if one assumes that the chemical potential in the tilted crystal is not set by the primary crystal and the two crystals have distinct chemical potentials and carrier densities. This implies that carriers cannot freely flow across the twinning plane, the interface between the two crystals.

\section{Angle-resolved Nernst effect}
We measured the Nernst effect with a standard one-heater-two-thermometer set-up.  The Nernst voltage was measured with two pairs of electrodes. As the magnetic field rotates, we switched from one pair of electrodes to another in order to measure the component of electric field in the plane perpendicular to the magnetic field. For each rotating plane, the experiment was performed in a dilution refrigerator inserted into a superconducting magnet for fields below 12 T and in a He$^{4}$ cryostat inserted in a resistive water-cooled magnet for fields between 12 T and 28 T.

Fig. 1 presents the raw Nernst data as a magnetic field is swept in the (trigonal, binary plane). The data for the two other planes are shown in the supplementary section. Each Nernst peak is associated with the intersection of a Landau level and the chemical potential. As seen in the bottom panel of the figure, different sets of data obtained at low and high magnetic field and with different pairs of electrodes match with each other. The peaks associated with the hole pocket can be clearly identified and are represented by red solid circles in the figure. But other peaks remain, whose origin is harder to pin down.

\section{Theoretical model}
For electrons at $L$-points, we employed the extended Dirac Hamiltonian introduced in ref. \cite{zhu2}. The parameters used here is also the same as those in ref. \cite{zhu2}, where the fitting is carried out below 12 T.

At high field, however, we have to take into account the interband coupling between the lowest Landau levels (LLLs) of the conduction and valence bands\cite{vecchi}, while it is negligible below 12 T. At zero field, the conduction and valence bands will not be coupled since they have opposite parity. On the other hand, at a finite field, the Bloch band picture is no more valid due to the complex interband effect of a magnetic field\cite{yf}, which can couple the two LLLs. Thus the coupling should be proportional to the field.
Such a interband coupling scenario actually agrees with the magneto-optical measurements\cite{vecchi}.

Here we introduce the modified LLL to take into account the interband coupling between the two LLLs in the form
\begin{align}
	E_{0, -}^{\rm inter}&=\pm \sqrt{
	\left\{\epsilon(k_{z}) -\frac{\tilde{g}'\beta_{0}}{4}B
	\right\}^{2}
	+\left(V \beta_{0} B \right)^{2}
	},\\
	\tilde{g}'&=g' \left(1+ V' \frac{|g'|\beta_{0}}{\Delta} B \right),
\end{align}
and $\epsilon (k_{z}) =\sqrt{\Delta^{2}+\hbar^{2}k_{z}^{2}\Delta/m_{z}}$, which are obtained by simplifying the model used in \cite{vecchi}.
($\beta_{0}=|e|\hbar / m_{\rm e}c=0.1158$ meV/T.)
Here, two new parameters are introduced: $V$ expresses the magnitude of the interband coupling, and $V'$ is the correction to the additional g-factor $g'$, both of which give contributions proportional to $B$.
Both $V$ and $V'$ are scalars and dimensionless, whereas the longitudinal mass $m_{z}$ and $g'$ are tensors as is given in ref. \cite{zhu2}.

The separation between the two LLLs decreases with increasing magnetic field in the low-field region;
the separation takes its minimum value at $B_{\rm min}$, and then increases in the high-field region.
We set the parameters as $V=0.15$ and $V'=-0.0625$.
With these parameters, $B_{\rm min}\sim$ 30 T, and the minimum value of the separation is about 1 meV (See the supplementary section for a discussion of the different parameters used here and in ref.\cite{vecchi}) .

For holes at $T$-point, we employed the same parabolic model and parameters used in ref. \cite{zhu2}, except for the value of the spin mass along the trigonal direction, $M_{\rm s 3}= 10000$. In the previous version of the model, this parameter was 200 (See TABLE III in ref. \cite{zhu2}).
This minor modification does not change the spectrum below 12 T at all, and gives better agreement at high fields.

\section{Crystal structure and twins}

The key point in elucidating the mystery of additional lines in the Landau spectrum resides in the rather exotic rhombohedral crystal structure of bismuth. The unit cell is obtained by pulling a cube along its body diagonal. This departure from higher symmetry is believed to be triggered by a Peierls transition\cite{peierls}. The distortion can be visualized by considering a symmetry-lowering operation on a tetrahedron. In absence of distortion, each atom would be at a vertex of a tetrahedron. At each of these vertices, the angle between two adjacent edges is 60 degrees (Fig. 2a). The tetrahedron remains invariant by a $2\pi/3 $ rotation around any of the four equivalent trigonal axes. The structural distortion corresponds to a displacement of one of the four vertices, which reduces the angle between the three adjacent edges at that particular vertex to less than 60 degrees. In this distorted structure, out of four original trigonal axes, only one survives and the structure keeps its threefold rotational symmetry around this trigonal symmetry.

Now, any of the four vertices of the tetrahedron can be subject to such a distortion (Fig. 2a). Therefore, bismuth crystals can have up to four domains. When at least two of such domains are simultaneously present in a crystal, the crystal is twinned. The atomic positions near a twin boundary in bismuth, originally documented by Hall\cite{hall}, is illustrated in Fig. 2b. A typical bismuth single crystal has many of such twin boundaries. Their presence can be used to establish a direction for trigonal and bisectrix axes and thus determine the sign of the tilt angle of the electron ellipsoid\cite{brown}. Twin boundaries have been directly detected by STM measurements\cite{edelman2}. When a sample is twinned, in addition to the principal crystal, there is a secondary crystal, which has its trigonal axis tilted by 108 degrees respective to the original crystal. The two crystals share one binary axis (fig. 2c).

\section{Landau spectrum}

Fig. 3 compares the theoretical prediction with experimental results for three planes of rotation. For each plane of rotation, a panel shows  both the Nernst peaks resolved by the experiment and the expected theoretical lines for holes and electrons. There is no ambiguity for hole peaks. As for electron peaks, we have selected Nernst peaks displaying an angular dependence close to what is theoretically expected. As seen in the figure, the agreement is excellent for holes, and rough for electrons. In the case of binary-bisectrix plane (Fig. 4c), the experimentally resolved hole peaks are sometimes split because of a small residual misalignment. When the field is oriented perpendicular to the trigonal axis, the Zeeman splitting of holes is extremely sensitive to the orientation and suddenly vanishes when the field is perpendicular to the trigonal axis (See panels a  and b of the same figure).

Given the quality of the agreement between theory and experiment in case of holes, it is natural to wonder about the origin and significance of the discrepancy observed in the case of electrons. The discrepancy is particularly visible at high fields. The theoretical lines for electrons tend to occur in magnetic fields significantly lower than the experimental ones. This is particularly the case for the angular window around the trigonal axis. We failed to find adequate theoretical parameters to close this gap between theory and experiment.

\section{The additional lines}

In Fig. 3, only those peaks which could be attributed either to electron or holes were shown. Fig. 4 display additional peaks (i.e. those which are clearly not associated with electrons) together with the hole peaks. The experimental peaks are compared with what is theoretically expected for holes of a secondary tilted crystal. This secondary crystal shares one binary axis with the original crystal, but has its trigonal axis tilted 108 degrees off, as depicted in Fig. 2c. As seen in the figure, the agreement is good. It is useful to carefully examine panel c. As seen in this figure, the additional lines merge with the hole lines when the field is aligned along one of the three binary. This is the binary axis common to the two twinned crystals. When the magnetic field is oriented along this binary axis, the spectrum does not show any additional features. In this configuration, the two crystals have the same crystal axis along the magnetic field and share an identical Landau spectrum.

Previous studies of the high-field phase diagram of bismuth focused on the angular window close to the trigonal axis\cite{luli,yang}. As seen in Fig. 5, in this range, the results obtained here are quite similar to those found in a previous study of angle-resolved Nernst effect\cite{yang} on another bismuth crystal. In both cases, experiment detects several lines and in particular three unexpected high-field peaks when the field is oriented along the trigonal axis as originally reported back in 2007\cite{behnia1}.  The additional peaks are not fractional peaks as initially assumed\cite{behnia1}, but can be indexed as $2_{h}^{-}$, $1_{h}^{+}$ and $3_{h}^{-}$ of a secondary tilted crystal as expected in the twinning scenario.

As seen in Figures 4 and 5, in the case of hole lines, the agreement between theory and experiment is impressive, both for the primary and the tilted crystal. This agreement has been obtained by assuming that the chemical potential continuously shifts with increasing magnetic field to preserve charge neutrality. Now, this field-induced shift in chemical potential strongly depends on the relative orientation of the magnetic field and the crystal axes\cite{zhu2}. One can think of two possibilities in presence of a strong magnetic field: i) the twinned crystals share a common chemical potential as they do at zero field. ii) The twinned crystals do not share a common chemical potential and as the magnetic field is applied, a difference between their Fermi energies and their carrier densities gradually builds up.

According to our results, the second scenario clearly prevails. If one assumes that the twinned crystals share a common chemical potential set by the primary crystal, then theory fails to reproduce the experimentally-resolved spectrum (See the supplementary section). On the other hand, the agreement between theory and experiment is satisfactory, if one assumes that the each of two crystals keeps its own chemical potential set by its local environment with respect to the orientation of the magnetic field. As seen in the lower panels of Fig. 5, this means that when the field is near the trigonal axis, there is a significant discontinuity both in chemical potential and in carrier density across the twin boundary.

\section{Conclusion and open questions}

In a typical bismuth crystal, twin boundaries abound but minority domains occupy a small fraction of the total volume of the sample\cite{twins}. It is therefore quite a surprise that the contribution of one of these domains is such that the magnitude of the Nernst peaks caused by their Landau levels is comparable to those of the primary crystal. What sets the magnitude of the Nernst quantum oscillations remains an open question. If the field-induced wrapping states (the 3D equivalent of the edge states of the Quantum Hall Effect\cite{koshino}), happen to play a role, then a twinned crystal with sufficient depth can generate a signal which weighs much larger than the volume it occupies.

The origin and the fine structure of the barrier  between the crystals with different chemical potential emerges as a new subject. Such barriers are common in semiconducting heterostructures. But the one under study has a number of intriguing peculiarities: the two crystals share the same structure and an identical chemical potential in zero magnetic field. Moreover, while the carrier density is different across the junction, the two systems are compensated and charge neutrality is locally preserved. Finally, the discontinuity in the chemical potential is a consequence of Landau quantification in a very anisotropic context. The wrapping states discussed by Koshino and co-workers\cite{koshino} have different Landau indexes across the twin boundary. Their length scale is set by the magnetic length. On the other hand, the length scale associated with charge reorganization at the interface is the Fermi wavelength.  In our context of investigation, these two length scales are comparable in size.

One conclusion of this study is that the one-particle theory can successfully reproduce the experimentally-resolved phase diagram of bismuth in a magnetic field as strong as 28 T. The fate of the three-dimensional electron gas in bismuth and in graphite diverge\cite{fauque3}. In graphite, the system undergoes a field-induced thermodynamic phase transition\cite{yaguchi}. This phase transition is believed to be a nesting-driven density-wave transition\cite{yoshioka}, which is expected occur due to the instability of the quasi-one-dimensional conductor emerging beyond the quantum limit. The absence of a phase transition in bismuth raises the following question: what protects bismuth from a similar instability?

\begin{acknowledgments}
We are grateful to J. Alicea, G. Mikitik, A. J. Millis,  Y. Sharlai and V. Oganesyan for fruitful discussions. This work is supported by Agence Nationale de Recherche as a part of QUANTHERM project and by EuroMagNET II under the EU contract number 228043. Y. F. is supported by MRL System of Osaka University, and also by Young Scientists (B) (No. 23740269) from Japan Society for the Promotion of Science.
\end{acknowledgments}


\begin{thebibliography}{10}
\bibitem{twins} According to our preliminary neutron diffraction measurements on the crystals used in this study,  the volume occupied by the major domain is an order of magnitude larger than each of the three minority domains.
\end{thebibliography}




\end{article}



\begin{figure}
\resizebox{!}{1\textwidth}{\includegraphics{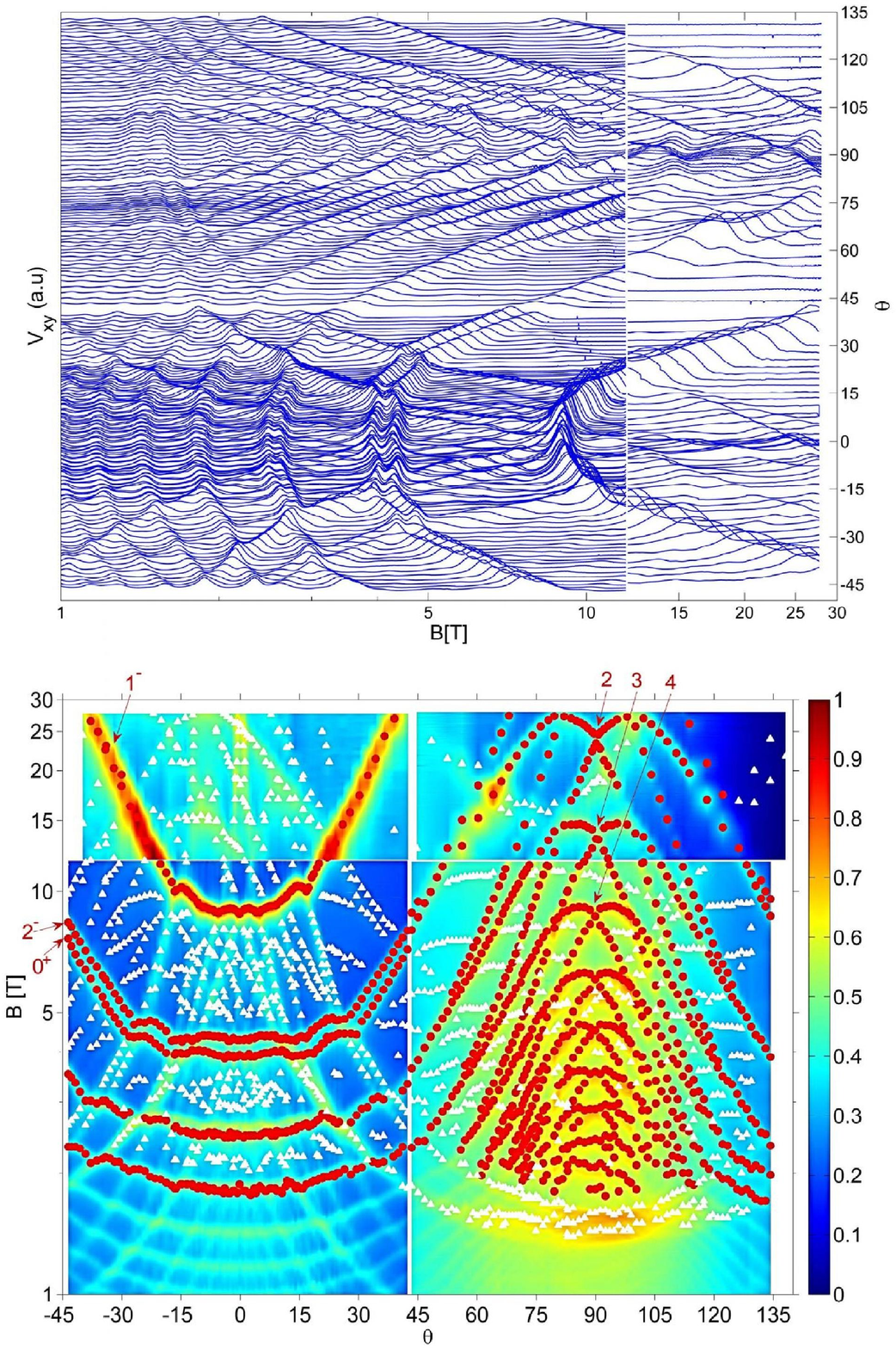}}

\caption{ \textbf{Top:} Nernst signal in a tilted magnetic field in the (trigonal, binary) plane for different tilt angles.\textbf{ Bottom:} Nernst peaks  superposed on a color plot of the same data. The angle $\theta$ represents the angle between the magnetic field and the trigonal axis. Those Nernst peaks, which are clearly associated with holes are represented by solid red circles and others by white triangles. The high- (low-)field data was obtained at 1.5 (0.49) K. Digits represent orbital indexes for hole Landau levels. They become spin-split ($\pm$) when $\theta\neq 90 ^{\circ}$.}
\end{figure}
\begin{figure}
\resizebox{!}{0.9\textwidth}{\includegraphics{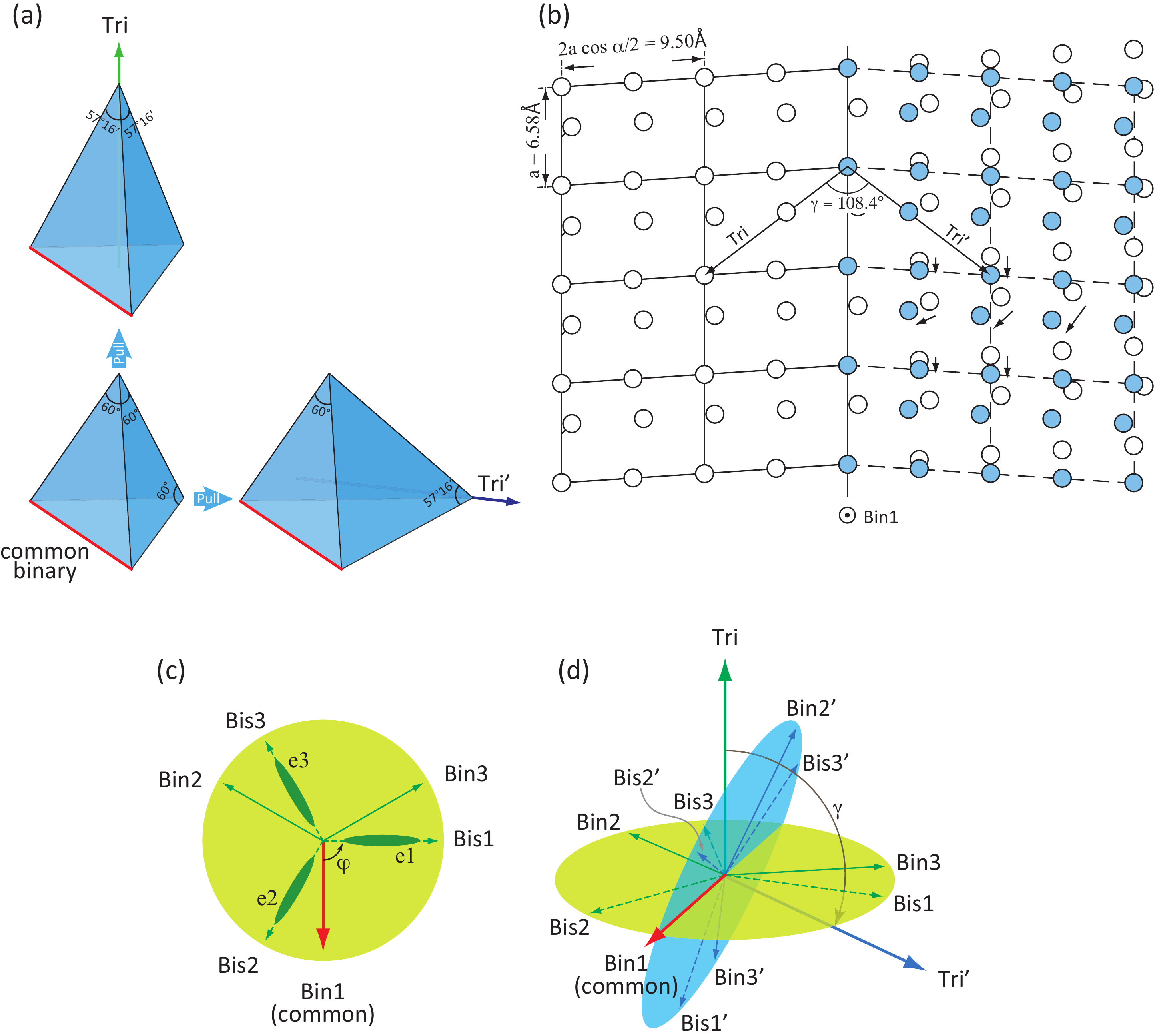}}
\caption{ \textbf{a) Structural distortion in bismuth: }The crystal structure of bismuth can be assimilated to a distorted tetrahedron. The tetrahedron can be distorted by pulling along any of the four different trigonal axes. At the corresponding vertex, the angle between each pair of adjacent edges becomes lower than 60 degrees. \textbf{b) Sketch of a twin boundary in bismuth after\cite{hall}}: Open circles represent atoms in the undistorted crystal. Solid circles represents atomic positions in the twinned crystal. The angle between the trigonal axes of the two crystals is about 108 degrees. The common binary axis for the two crystals is perpendicular to the plane of the figure. \textbf{c) Conventions used in this text:} The green plane and axis are the original trigonal plane and axis. The blue ones are tilted planes and axes. The two structures share one binary axis.}
\end{figure}

\begin{figure}
\resizebox{!}{1\textwidth}{\includegraphics{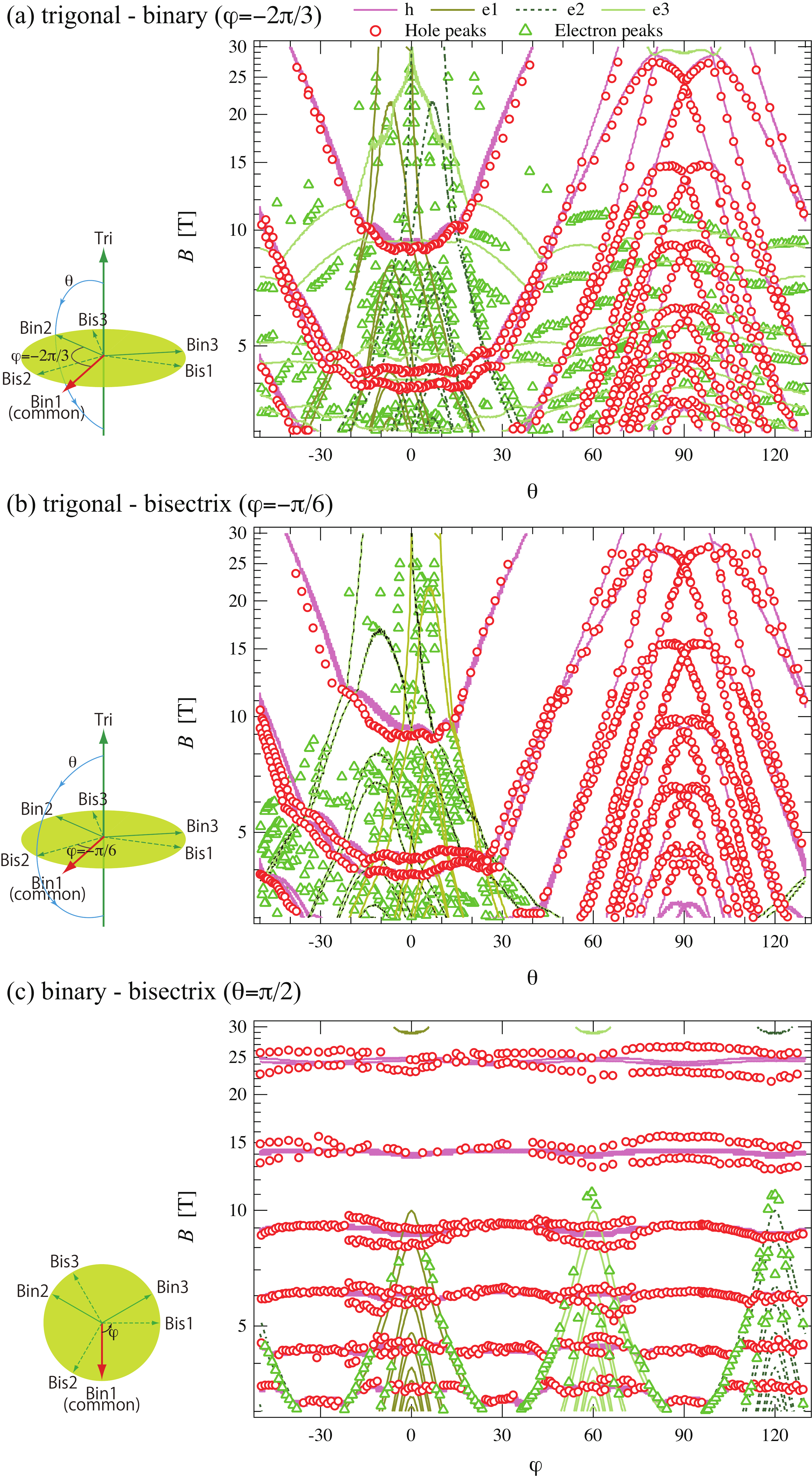}}
\caption{ The Landau spectrum for holes and electrons according to theory (solid lines) and experiment (symbols). The three panels correspond to the three rotating planes. Peaks resolved by experiment and identified as arising from holes (red open circles) or electrons (green open triangles) are displayed.}
\end{figure}

\begin{figure}
\resizebox{!}{1\textwidth}{\includegraphics{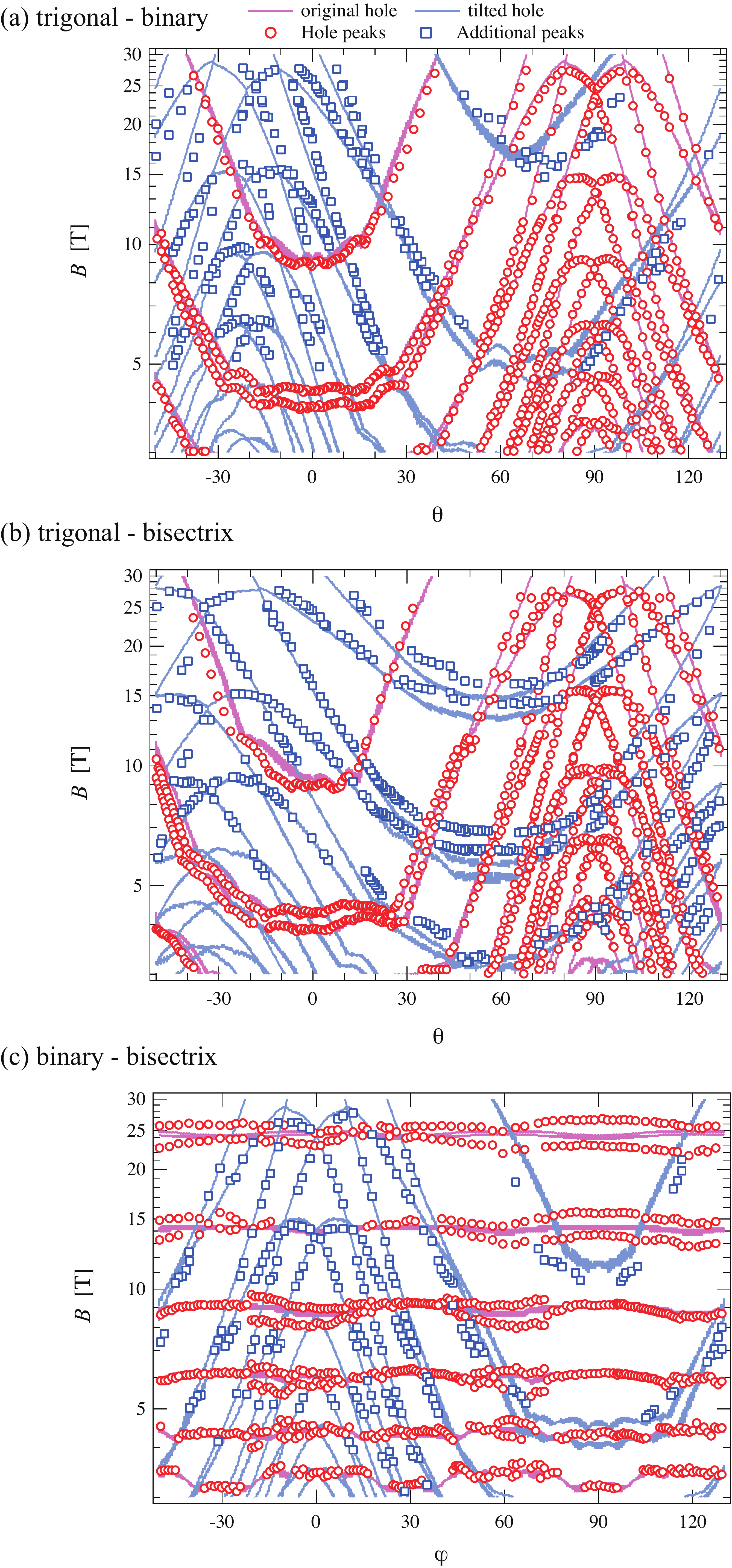}}
\caption{ The Landau spectrum for holes in the original and tilted crystal according to theory (solid lines) and experiment (symbols). The three panels correspond to the three rotating planes.  Additional peaks resolved by the experiment are represented by blue open squares.}
\end{figure}

\begin{figure}
\resizebox{!}{0.75\textwidth}{\includegraphics{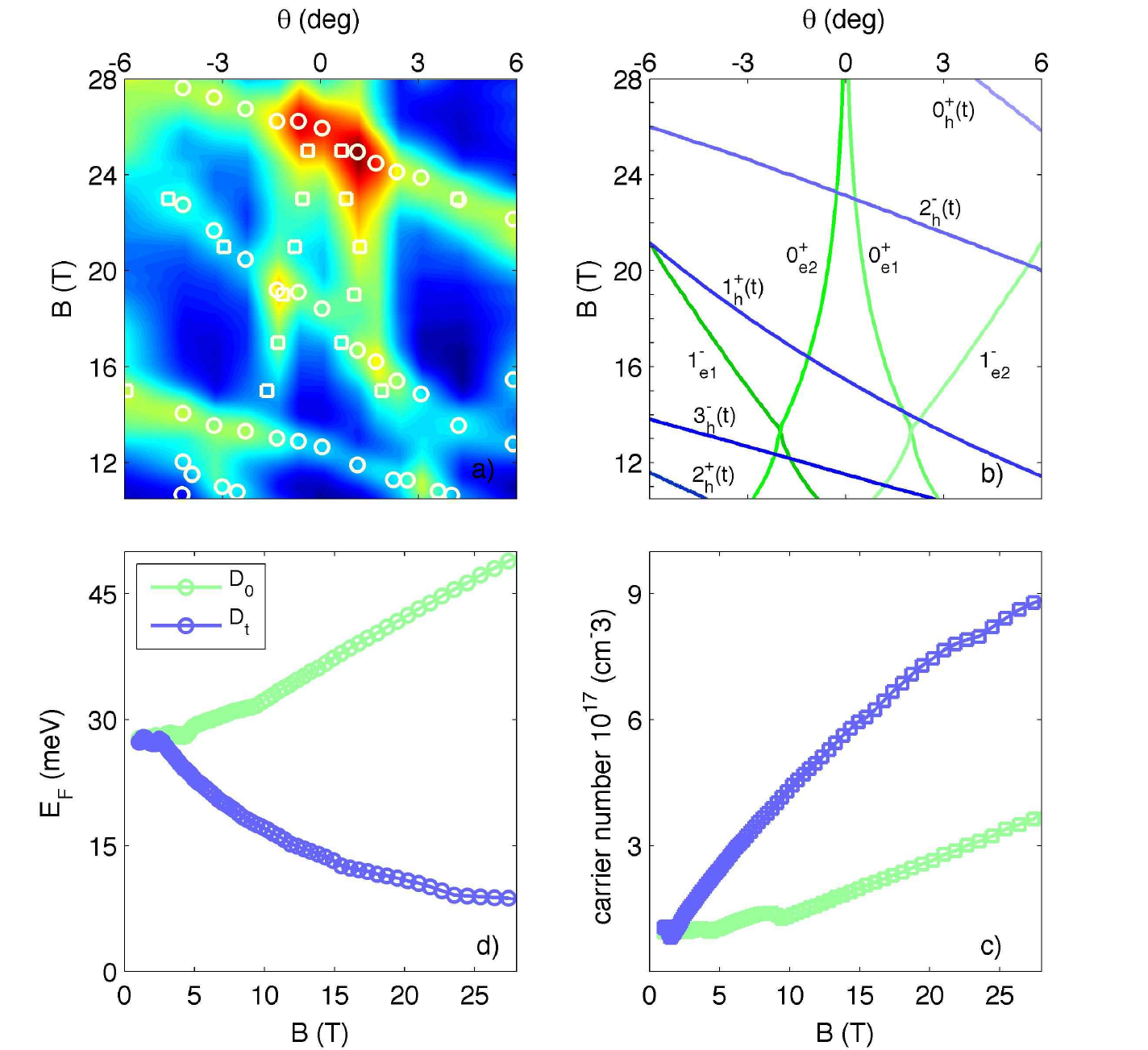}}
\caption{Top:  Experimental (a) and theoretical (b) Landau spectrum for field oriented close to the trigonal axis.  The lines are indexed according to their orbital and spin quantum number. The additional lines correspond to tilted hole lines. Bottom: The evolution of the carrier density (c) and chemical potential (d) for the primary (D$_{0}$) and tilted(D$_{t}$) crystals when the magnetic field is parallel to the trigonal axis. }
\end{figure}







\section*{Supplementary material}

# Supplementary information for "Landau spectrum of bismuth in extreme quantum limit: field-induced mismatch in chemical potential across twin boundaries"

Zengwei Zhu[1], Benoît Fauqué[1], Liam Malone[2], A. B. Antunes[2], Yuki Fuseya[3] and Kamran Behnia[1]

(1)LPEM (UPMC-CNRS), ESPCI, 75005 Paris, France

(2) LCMI (CNRS), 38042 Grenoble , France

(3) Department of Materials Engineering Science,
Osaka University, Toyonaka, Osaka 560-8531, Japan

# I. EXPERIMENTAL DETAILS

We measured the Nernst effect with a standard one-heater-two-thermometer set-up, which was rotated in a magnetic field using Attocube piezoelectric rotators. All three perpendicular rotating planes were scanned in order to extract the map for the whole solid angle. In the main text, the raw data were presented for one plane of rotation. We present in Fig. 1 and Fig. 2, the data for two other planes of rotation.

# II. THEORETICAL DETAILS

The model used here is slightly different from the one used in ref.[1]. The magnetic field dependence of the lowest Landau levels (LLLs) for the present parameters ($V = 0.15$ and $V' = -0.0625$) are shown for $B \parallel$ binary in Fig. 3 (a). The origin of the energy is at the center of the band gap, while in Fig. 3 of ref. [1], it is taken at the bottom of the conduction band.

The current model takes into account an interband effect induced by magnetic field, which can couple the two Lowest Landau levels (LLLs). In the present case, the separation between the two LLLs takes its smallest value at $B_{\mathrm{min}} \sim 30$ T, and the LLL in the conduction band does not cross the Fermi energy up to 45 T.

The result with $V = 1.26$ and $V' = 0.253$, which are derived from the parameters used in ref. [2], is shown in Fig. 3 (b). The result agrees well with Fig. 2 of ref. [2], even though our model is slightly different from theirs. (Note that the result shown in Fig. 2 of ref. [2] is not the result obtained experimentally, but is the result obtained theoretically.) In their case, the separation takes its smallest value at $B_{\mathrm{min}} \sim 10$ T, and rapidly increase above 10 T, due to the large values of $V$ and $V'$.

As seen in Fig. 3 (b), the LLL in the conduction band crosses the Fermi energy at around 20 T, and the electron pocket becomes completely empty above this field. This crossing would have generated an additional peak. However, such a peak has not been observed in our Nernst experiments up to 28 T. Therefore, the current experimental results favor the situation sketched in Fig. 3 (a).

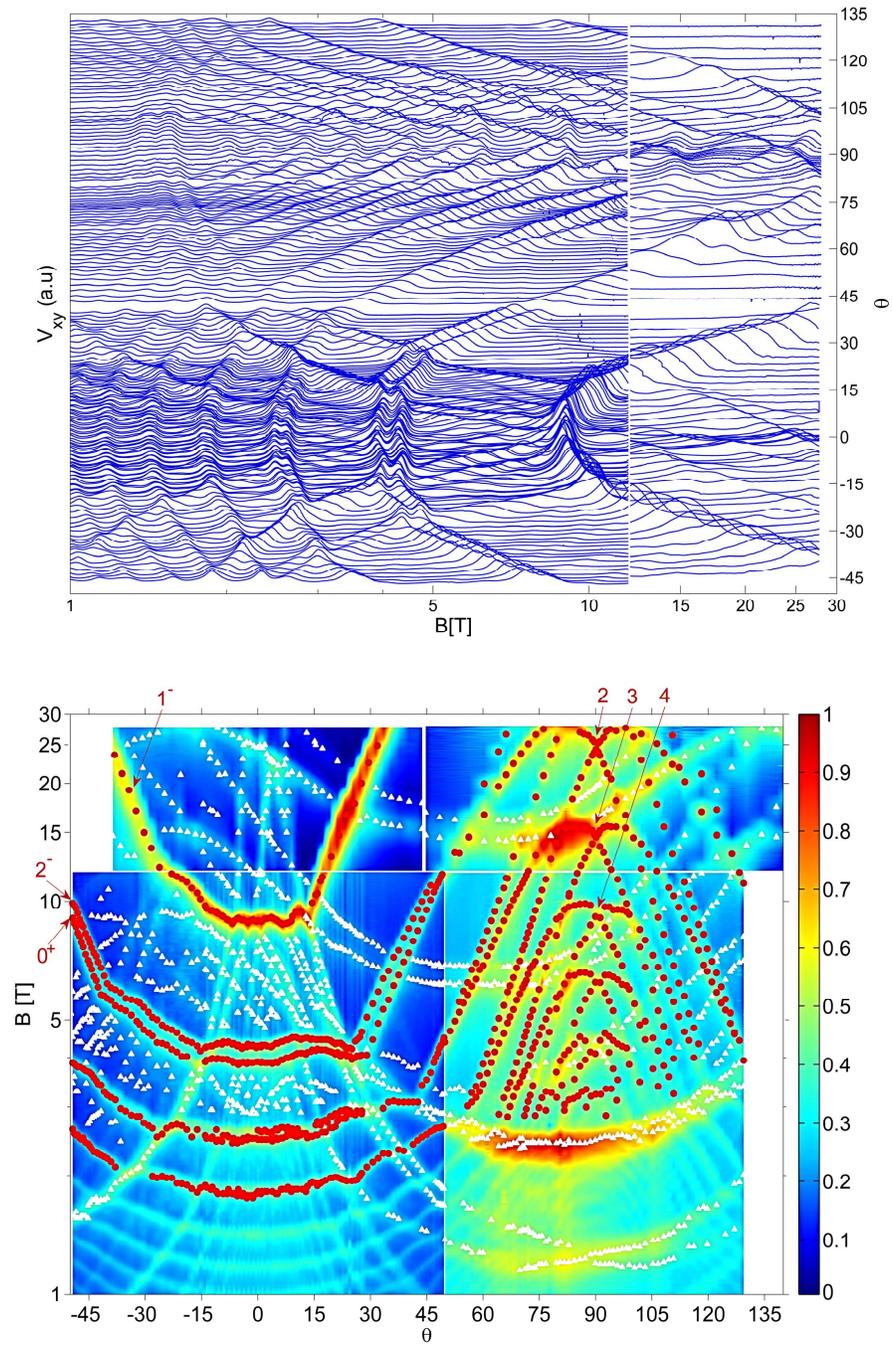

FIG. 1: **Supplement:**Top: Nernst signal in a tilted magnetic field in the (trigonal, bisectrix) plane for different tilt angles. Bottom: Nernst peaks superposed on a color plot of the same data.

3

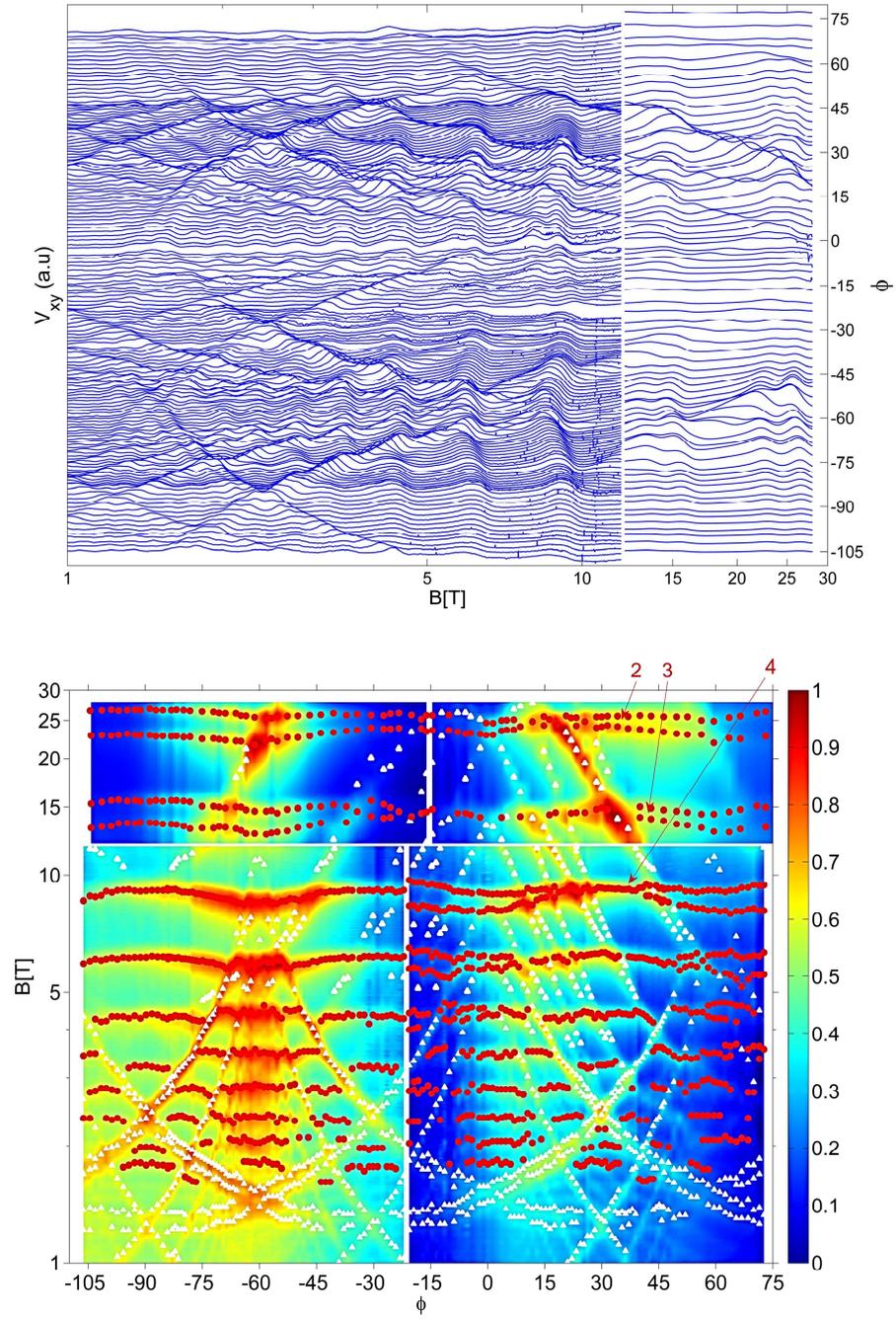

FIG. 2: **Supplement:** Top: Nernst signal in a tilted magnetic field in the (binary, bisectrix) plane for different tilt angles. Bottom: Nernst peaks superposed on a color plot of the same data.

4

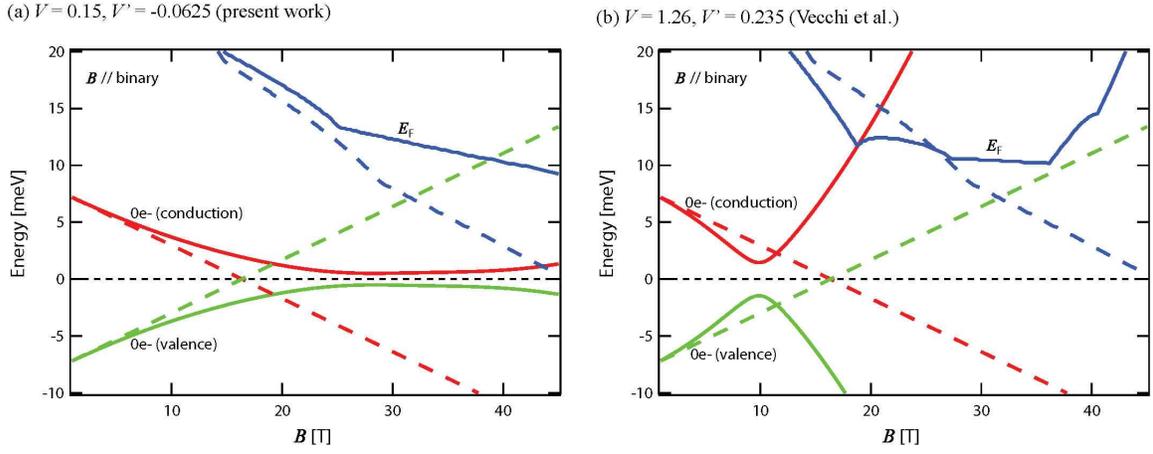

FIG. 3: **Supplement:** Magnetic field ($B \parallel$ binary) dependence of the lowest Landau levels (0e-) and the Fermi energy ($E_\mathrm{F}$) for (a) $V = 0.15$ and $V' = -0.0625$ (present work), and (b) $V = 1.26$ and $V' = 0.253$ (ref. [2]). The dashed lines are the results without considering the interband coupling between the two lowest Landau levels.

## III. FAILURE OF THE COMMON CHEMICAL POTENTIAL SCENARIO

In the manuscript, we have compared theory and experiment in the case of different chemical potentials for the primary and tilted crystals (Fig. 4 and Fig. 5 in the main manuscript). In Fig. S4, we show the results of theoretical calculations for one rotating plane assuming a common chemical potential for both crystals. As seen in the figure, there is a significant discrepancy between theory and experiment. The failure of theory by a very large margin in this scenario compared to its success when the chemical potential assumed to be different backs one of our main conclusions.

---

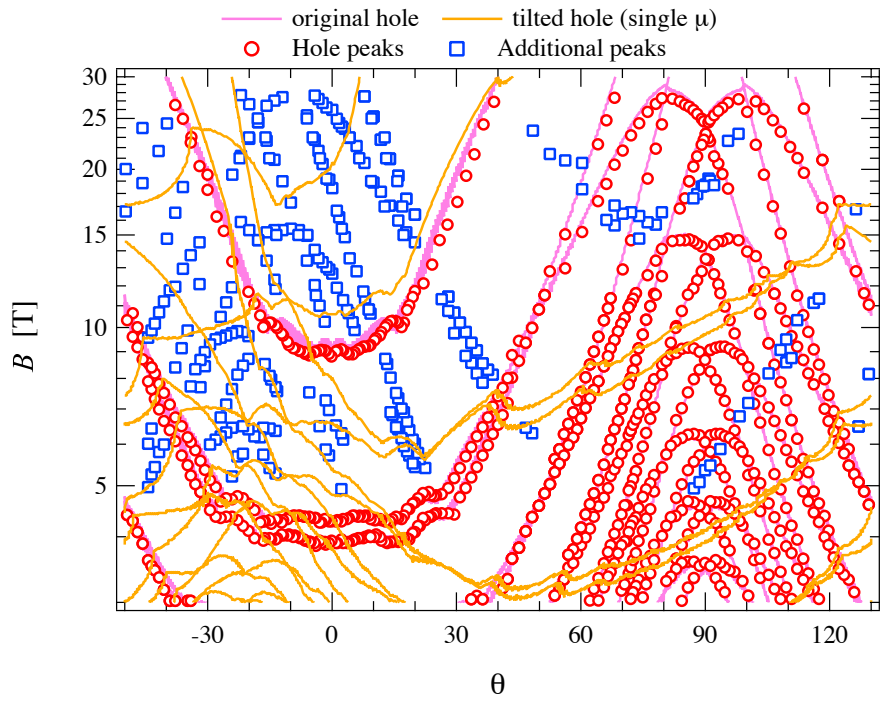

FIG. 4: **Supplement:** The Landau spectrum for holes in the original and tilted crystal according to theory (solid lines) and experiment (symbols) for the (trogonal, bisectrix in case of a single chemical potential. Additional peaks resolved by the experiment are shown by blue empty squares. They do not match the theoretical lines.


\end{document}